\documentclass[useAMS,usenatbib,fleqn]{mn2e}
\usepackage{amsmath}
\usepackage{graphicx}
\usepackage{color}
\usepackage{xspace}
\usepackage{url}
\usepackage{amssymb}
\usepackage{mathtools} 
\usepackage{enumitem}
\usepackage{natbib}

\def \aap{A\&A}

\def \apjl{ApJ}
\def \apjs{ApJS}

\newcommand{\ias}{{\sc \tt IAS15}\xspace}

\newcommand{\mer}{{\sc \tt MERCURY}\xspace}
\newcommand{\mercury}{{\sc \tt MERCURY}\xspace}
\newcommand{\mercurius}{{\sc \tt MERCURIUS}\xspace}

\title{Hybrid Symplectic Integrators for Planetary Dynamics}
\date{Accepted for publication by MNRAS on 12th March 2019. Submitted on 19th January 2019.}

\author[Rein et.al.]{Hanno Rein$^{1,2}$\thanks{E-mail: \texttt{hanno.rein@utoronto.ca}}, 
    David M. Hernandez$^{3,4}$, 
    Daniel Tamayo$^{5}$\thanks{NHFP Sagan Fellow},
    Garett Brown$^{1,6}$, 
    \newauthor 
    Emily Eckels$^{7}$, 
    Emma Holmes$^{8}$, 
    Michelle Lau$^{9}$,
    R\'ejean Leblanc$^{1,6}$,
    Ari Silburt$^{1,2}$\\ 
$^1$ Department of Physical and Environmental Sciences, University of Toronto at Scarborough, Toronto, Ontario M1C 1A4, Canada\\
$^2$ Department of Astronomy and Astrophysics, University of Toronto, Toronto, Ontario, M5S 3H4, Canada\\
$^3$ {Harvard--Smithsonian Center for Astrophysics, 60 Garden St., MS 51, Cambridge, MA 02138, USA} \\
$^4$ {MIT Kavli Institute for Astrophysics and Space Research, 77 Massachusetts Ave., Cambridge, MA 02139, United States}\\
$^5$ {Department of Astrophysical Sciences, Princeton University, Princeton, New Jersey 08544, United States}\\
$^6$ Department of Physics, University of Toronto, Toronto, Ontario, M5S 3H4, Canada\\
$^7$ Department of Mathematics, Emory University, 201 Dowman Drive, Atlanta, GA 30322, USA\\
$^8$ Department of Mathematics and Statistics, McMaster University, 1280 Main St. W., Hamilton, Ontario, L8S 4K1, Canada\\
$^9$ Department of Physics, Imperial College London, London, SW7 2AZ, United Kingdom\\
}

\vspace{0.5\baselineskip}

\begin{document}
\maketitle

\begin{abstract}
Hybrid symplectic integrators such as \mer are widely used to simulate complex dynamical phenomena in planetary dynamics that could otherwise not be investigated.
A hybrid integrator achieves high accuracy during close encounters by using a high order integration scheme for the duration of the encounter while otherwise using a standard 2nd order Wisdom-Holman scheme, thereby optimizing both speed and accuracy.
In this paper we reassess the criteria for choosing the switching function that determines which parts of the Hamiltonian are integrated with the high order integrator.
We show that the original motivation for choosing a polynomial switching function in \mer is not correct.
We explain the nevertheless excellent performance of the \mer integrator and then explore a wide range of different switching functions including an infinitely differentiable function and a Heaviside function.
We find that using a Heaviside function leads to a significantly simpler scheme compared to \mer, while maintaining the same accuracy in short term simulations.
\end{abstract}

\begin{keywords}
methods: numerical --- gravitation --- planets and satellites: dynamical evolution and stability 
\end{keywords}

\section{Introduction}
\label{sec:intro}
Since ancient times astronomers have predicted the locations of planets in the night sky.
But only the advent of computers has made it possible to calculate the orbital evolution of planetary systems over millions and even billions of years \citep{LaskarGastineau2009}.
One major breakthrough was the development of mixed variable symplectic maps for the $N$-body problem which we will refer to as Wisdom-Holman integrators \citep{WisdomHolman1992}. 
The idea behind the Wisdom-Holman integrator is to split the motion of a planet into two steps: a dominant Keplerian motion around the star (the Kepler step), and the motion due to small interactions between the planets (the interaction step). 

The Wisdom-Holman scheme works well as long as the interactions between planets can be considered perturbations to their predominant Keplerian motion around the star.
However, if two planets come close to each other, the planet-planet interactions can become the dominant part of the motion.
In that case the Wisdom-Holman integrator becomes inaccurate and might require excessively small timesteps to achieve an acceptable solution.

\cite{Duncan1998} and \cite{Chambers1999} provide a solution to this problem in the form of hybrid symplectic integrators.
The idea is to move terms from the interaction part to the Keplerian part during close encounters, to ensure that one part always remains a perturbation. 
Although solving the Keplerian part during a close encounter now becomes more difficult than simply solving Kepler's equation, it can be done relatively efficiently with a high order integrator such as a Burlish-Stoer or Gau\ss-Radau scheme.
As long as close encounters happen infrequently, the high order integrator is rarely used  and has a negligible effect on the runtime.
The scheme is therefore almost as fast as a Wisdom-Holman integrator.
Furthermore, the scheme is also symplectic, ensuring good long term conservation of energy and angular momentum.

In this paper we look specifically at the \mer integrator.
\mer is freely available online and has become a widely used tool in running many otherwise impossible types of simulations. 
We discuss inconsistencies in the original paper by \cite{Chambers1999} which describes the \mer algorithm\footnote{A similar discussion in \cite{Duncan1998} is correct, but it describes a slightly different and more complex scheme.}.
We note that \cite{Wisdom2017} also looked at these inconsistencies and came independently of us to the same conclusion.
Despite these inconsistencies, the resulting scheme has very good characteristics.
We provide mathematical and physical explanations for them in this paper.
With this new understanding, it is easy to construct new hybrid integrators. 
We describe one new integrator that is much simpler than \mer, but achieves the same accuracy.

\section{Hybrid Symplectic Integrators}
We begin by introducing the notation and coordinate system we use. 
We have one star and an additional $N$ planetary bodies with Cartesian coordinates $\mathbf q_i$ and canonical momenta $\mathbf p_i$ in some inertial frame.
We will identify the stellar object with $i=0$ and the planets with $i>0$.
The $N$-body Hamiltonian in these coordinates is
\begin{eqnarray}
    H &=& \sum_{i=0}^{N} \frac{p_i^2}{2m_i} - G \sum_{i=0}^{N} \sum_{j=i+1}^{N} \frac{m_i m_j}{\left|\mathbf q_i-\mathbf q_j\right|}.\label{eq:nbody}
\end{eqnarray}
To evolve this system forward in time we could calculate the equations of motion for this Hamiltonian. 
However, in general the resulting differential equations cannot be solved analytically and are usually stiff and thus hard to solve numerically.

The hybrid integration schemes discussed in this paper make use of democratic heliocentric coordinates $Q_i$ and corresponding canonical momenta $P_i$ (see Appendix~\ref{appendix:coord}).
The advantage of these coordinates is that the Hamiltonian can be written as a sum of four terms:
\begin{eqnarray}
     H &=& H_0 + H_{\rm K} + H_{\rm I}+ H_{\rm J}, \label{eq:mvs}
\end{eqnarray}
where 
\begin{eqnarray}
    H_{0}  &=&  \frac{P_0^2}{2M} \label{eq:mvs0}\\
    H_{K} &=& \sum_{i=1} \left( \frac{P_i^2}{2m_i} - \frac{Gm_0m_i}{Q_i} \right)\nonumber\\
           && -G \sum_{i=1}\sum_{j=i+1} \frac{m_i m_j}{Q_{ij}} \left(1-K(Q_{ij})\right)\\
    H_{I}  &=& -G \sum_{i=1}\sum_{j=i+1} \frac{m_i m_j}{Q_{ij}} K(Q_{ij})\\
    H_{J}  &=& \frac1{2m_0} \left( \sum_{i=1} \mathbf P_i \right)^2. \label{eq:mvsJ}
\end{eqnarray}
In the above splitting, we've introduced the arbitrary scalar function $K(r)$ which depends only on the distance between pairs of particles.
Note that the terms involving $K$ cancel out when adding $H_K$ and $H_I$ and thus do not change the evolution of the system. 

In this paper, we will use the letter $K$ to label switching functions that appear in the Hamiltonian and refer to them as \textit{Hamiltonian switching functions}. 
Later, we will also encounter \textit{force switching functions} that first appear in the equations of motions.
To avoid any confusion, we will use the letter $L$ to label force switching functions.
There is a relation between $K$ and $L$ which we derive in Sec.~\ref{sec:mer}.

Each of the above terms has a physical interpretation.
$H_0$~describes the motion of the centre of mass. 
The term~$H_K$ describes the Keplerian motion of the planets, ignoring  planet-planet interactions (while $K=1$).
This term is separable and can be solved for each planet independently.
The term~$H_I$ corresponds to the planet-planet interactions.
The term~$H_J$ is related to the barycentric motion of the star and is referred to as the jump term (it changes the positions of particles instantaneously but keeps the momenta constant).

We can calculate the equations of motion for each of these Hamiltonians.
The resulting differential equations for $H_0$, $H_I$, and $H_J$ are trivial to solve.
Only the equations for $H_K$ require more work, i.e. solving Kepler's equation (while $K=1$).
Evolving the system under the influence of a Hamiltonian can be thought of as an operator acting on a state corresponding to the initial conditions.
The idea of an operator splitting integration method is to approximate the evolution of the full system by consecutively evolving the system under the partial Hamiltonians. 
We can construct arbitrarily high order integrators by applying these operators in the right order for different amounts of time \citep{Yoshida1990}.
In particular, a second-order integrator can be constructed from the above splitting by using the following chain of operators 
\begin{eqnarray}
    \widehat H_{WH}(dt) &\equiv& 
    \widehat H_{I}(dt/2) \circ  
    \widehat H_J(dt/2) \circ  \label{eq:whsplit}
    \widehat H_0(dt)  \nonumber \\
    && \circ \, 
    \widehat H_{K}(dt) \circ  
    \widehat H_J(dt/2) \circ  
    \widehat H_{I}(dt/2), 
\end{eqnarray}
where an operator $\widehat H_A(dt)$ corresponds to evolving the system under the Hamiltonian $H_A$ for a time $dt$.
Note that we recover the standard Wisdom-Holman integrator in democratic heliocentric coordinates if we set $K=1$.

For any finite timestep $dt$, $\widehat H_{WH}$ is only approximately equal to $\widehat H$.
We can analyze the error of this splitting scheme with the help of the Baker-Campbell-Hausdorff (BCH) formula.
The splitting scheme error arises because some pairs of operators do not commute.
The error consists of an infinite series of commutators between operators.
In our case, the action of an operator corresponds to solving the differential equations resulting from its respective Hamiltonian.
As a result, the splitting scheme error can be expressed as a series of nested Poisson brackets \citep[see e.g.][]{HernandezDehnen2016}.
One caveat to keep in mind when working with the BCH formula is that it is only a formal asymptotic series and does not necessarily converge everywhere in phase space.
We come back to this issue later.

The commutators that appear in the BCH formula can be grouped together by powers of the timestep $dt$.
The specific splitting scheme in Eq.~(\ref{eq:whsplit}) is time symmetric and therefore only even orders of $dt$ appear \citep{Yoshida1990}. 
Further note that $\widehat H_J$ commutes with $\widehat H_I$.
Thus, the lowest order non-zero error terms that appear are:
\begin{eqnarray}
    && \frac1{12} \small[\small[\widehat H_J,\widehat H_K\small],\widehat H_K\small],\quad \frac{1}{24}\small[\small[\widehat H_J,\widehat H_K\small],\widehat H_J\small],\nonumber\\
    && \frac1{12} \small[\small[\widehat H_I,\widehat H_K\small],\widehat H_K\small], \quad  \frac1{24}\small[\small[\widehat H_I,\widehat H_K\small],\widehat H_I\small],\nonumber\\
    && \text{and}\quad \frac1{12} \small[\small[\widehat H_J,\widehat H_K\small],\widehat H_I\small].
\end{eqnarray}
In this paper, we do not attempt to calculate approximations to these expressions analytically using Poisson brackets, but do so numerically. 
This allows us to easily extend the calculations to arbitrarily high order (7th order in our case). 

There are two way to approach this. 
Consider a commutator of the form $[\widehat A, \widehat B]$.
Starting from a state $z$, we can calculate $\delta z \equiv (\widehat A \widehat B) \, z - (\widehat B \widehat A)\, z $, and similarly for higher order commutators. 
If $\delta z=0$, then the commutators commute. 
To quantify the statement, we can use any norm on $\delta z$.
The measure we are typically interested in is the energy, or more specifically the energy error.
Note that calculating the energy of $\delta z$ does not make sense.
However, we can calculate the energy error by comparing the energy of the state $z$ to that of the state $z+\delta z$.
Alternatively, we can  calculate the state $z' \equiv  (\widehat B^{-1} \widehat A^{-1} \widehat B \widehat A)\, z$. 
Here, $z'$ is a valid state and we can directly compare the energy of the states $z$ and $z'$. 

In this paper we use the latter approach.
Note that it is a simple chain of operators and no additions or subtractions of state vectors are needed.
This approach avoids some issues with finite floating point precision compared to the standard approach where one needs to subtract nearly equal numbers multiple times when calculating the energy difference of the states $z$ and $z+\delta z$.
As a specific example, consider the first (third order) commutator that appears in the BCH formula of the hybrid symplectic intergators: $ \small[\small[\widehat H_J,\widehat H_K\small],\widehat H_K\small]$.
The effect of this commutator can be calculated by the following chain of operators:
\begin{eqnarray}
  &&
    \underbrace{
    \widehat H_{K}(-dt) \circ
    \widehat H_{J}(-dt) \circ  
    \widehat H_{K}(dt) \circ  
    \widehat H_{J}(dt)  }_{\mathrel{\widehat{=}} \small[\widehat H_J,\widehat H_K\small]^{-1}(-dt)}\circ  
    \widehat H_{K}(-dt) \nonumber\\  && \;\circ\; 
\underbrace{
     \widehat H_{J}(-dt) \circ 
    \widehat H_{K}(-dt) \circ  
    \widehat H_{J}(dt) \circ  
    \widehat H_{K}(dt)}_{\mathrel{\widehat{=}} \small[\widehat H_J,\widehat H_K\small](dt)}\circ  
    \widehat H_{K}(dt) \nonumber .
\end{eqnarray}

The two approaches are not exactly equal, only to within order $dt$.
Our approach is only an approximation of the commutator appearing in the BCH formula and is accurate to within $\mathcal{O}(dt)$ only.  
We could apply the Zassenhaus formula (the dual of the BCH formula) to calculate higher order corrections, which are again chains of the same operators. 
However, the approximations are good enough for our purpose: they will allow us to demonstrate which are the dominant contributions to the error terms in the BCH formula and understand their origins, especially for higher order commutators which are otherwise not easily tractable.

\section{The MERCURY Code}
\label{sec:mer}
The Wisdom-Holman map (setting $K=1$ in the above notation) performs poorly during close encounters because some of the commutators in the error term become large.
To construct an integrator that can more accurately resolve close encounters, \cite{Chambers1999} suggests to use a non-constant Hamiltonian splitting function $K$ as a basis for the \mer scheme.

The idea, as outlined in \cite{Chambers1999}, is that the switching function keeps the interaction Hamiltonian small even during close encounters between planets.
As a result the switching scheme errors from the BCH formula should remain small as well.

Let us use Hamilton's equations and derive the equations of motion for the Hamiltonian $H_{K}$:
\begin{eqnarray}
    \mathbf{  \dot Q}_i &=& \mathbf V_i\\
    \mathbf{  \dot V}_i &=& - \frac{Gm_0}{Q_i^3} \mathbf Q_i \nonumber\\
             && - G \sum_{\substack{j=1\\j\neq i}} \frac{ m_j}{Q_{ij}^3} \mathbf Q_{ij} \left(1-K(Q_{ij}) + Q_{ij}  K'(Q_{ij})\right), \label{eq:eomhkh}
\end{eqnarray}
where we have introduced the democratic heliocentric velocities
\begin{eqnarray}
    \mathbf{V}_i \equiv \frac{\mathbf P_i}{m_i}
    \quad \quad \quad
    \text{and therefore}
    \quad \quad \quad
    \mathbf{ \dot V}_i = \frac {\mathbf{ \dot P}_i}{m_i}.
\end{eqnarray}
We also used the fact that the gradient of $K$ satisfies
\begin{eqnarray}
    \frac{\partial K(Q_{ij})}{\partial Q_{ij}} = \left.\frac{\mathbf Q_{ij}}{Q_{ij}}\, \frac{\partial K(r)}{\partial r} \right|_{r=Q_{ij}}
    = \frac{\mathbf Q_{ij}}{Q_{ij}}\, K'(Q_{ij}).
\end{eqnarray}
The equations of motion above are not those of Keplerian motion but now also include interaction terms. 
Nevertheless they can be solved with a high order integrator such as a Bulirsch-Stoer algorithm or a high order Gau\ss-Radau integrator \citep{Everhart1985,ReinSpiegel2015}.

Similarly, the equations of motion for $H_{I}$ can be derived as
\begin{eqnarray}
    \mathbf{  \dot Q}_i &=& 0\\
    \mathbf{\dot V}_i &=&  - G \sum_{\substack{j=1\\j\neq i}} \frac{ m_j}{Q_{ij}^3} \mathbf Q_{ij} \left( K(Q_{ij}) 
             -  Q_{ij} K'(Q_{ij}) \right) \label{eq:eomhik}
\end{eqnarray}
These equations remain trivial to solve even in the case of $K\neq1$.

Somewhat surprisingly, the equations of motion derived above are not the equations of motion implemented in the \mer code. 
Specifically, the terms involving $K'(Q_{ij})$ are not present in the publicly available version of \mer. 
Therefore, the \mer code does also not correspond to the Hamiltonian splitting in Eq.~(\ref{eq:mvs}).

This then begs the question\footnote{
    This inconsistency was discussed at the Toronto Meeting on Numerical Integration Methods in Planetary Science in 2017 and an explanation was first published by \cite{Wisdom2017}.

We independently came to the same conclusion.
}: What \textit{is} \mer solving?
To find the answer to this question, we define a new switching function 
\begin{eqnarray}
    L(r) &\equiv& K(r) - r K'(r).  \label{eq:LK}
\end{eqnarray}
With this definition, we can rewrite the equations of motion in terms of $L$. 
For example, Eq.~(\ref{eq:eomhkh}) becomes
\begin{eqnarray}
    \mathbf{  \dot V}_i &=& - \frac{Gm_0}{Q_i^3} \mathbf Q_i - G \sum_{\substack{j=1\\j\neq i}} \frac{ m_j}{Q_{ij}^3} \mathbf Q_{ij} \left(1-L(Q_{ij}) \right).
\end{eqnarray}
Note that no derivatives of $L$ appear in the equations of motion.
We can think of $L$, which is changing between~$0$ and~$1$, as a weight in the inter-particle forces. 
We therefor refer to it as the force switching function.
If we set $L$ to the piecewise polynomial function
\begin{eqnarray}
    L_{\rm merc}(r) = \begin{cases} 0 & y\leq0\\
        10 y^3 - 15 y^4 + 6 y^5 & 0<y<1\\
        1 & y \geq 1,
    \end{cases}\label{eq:L}
\end{eqnarray}
with
\begin{eqnarray}
    y = \frac{r-0.1r_{\rm crit}}{0.9 r_{\rm crit}} \label{eq:y}
\end{eqnarray}
then we recover exactly the integration scheme as implemented in the \mer code\footnote{Note that this is not the function described in the paper which has roughly the same shape but is slightly different from the one implemented in the publicly available version of \mer. 
We here use the function implemented in the \mer code.}.
In the above equation, $r_{\rm crit}$ is some critical switching distance. 
The value depends on the specific application but it typically a small multiple of the mutual Hill radius of the particles.

\begin{figure}
    \centering
    \resizebox{0.99\columnwidth}{!}{\includegraphics{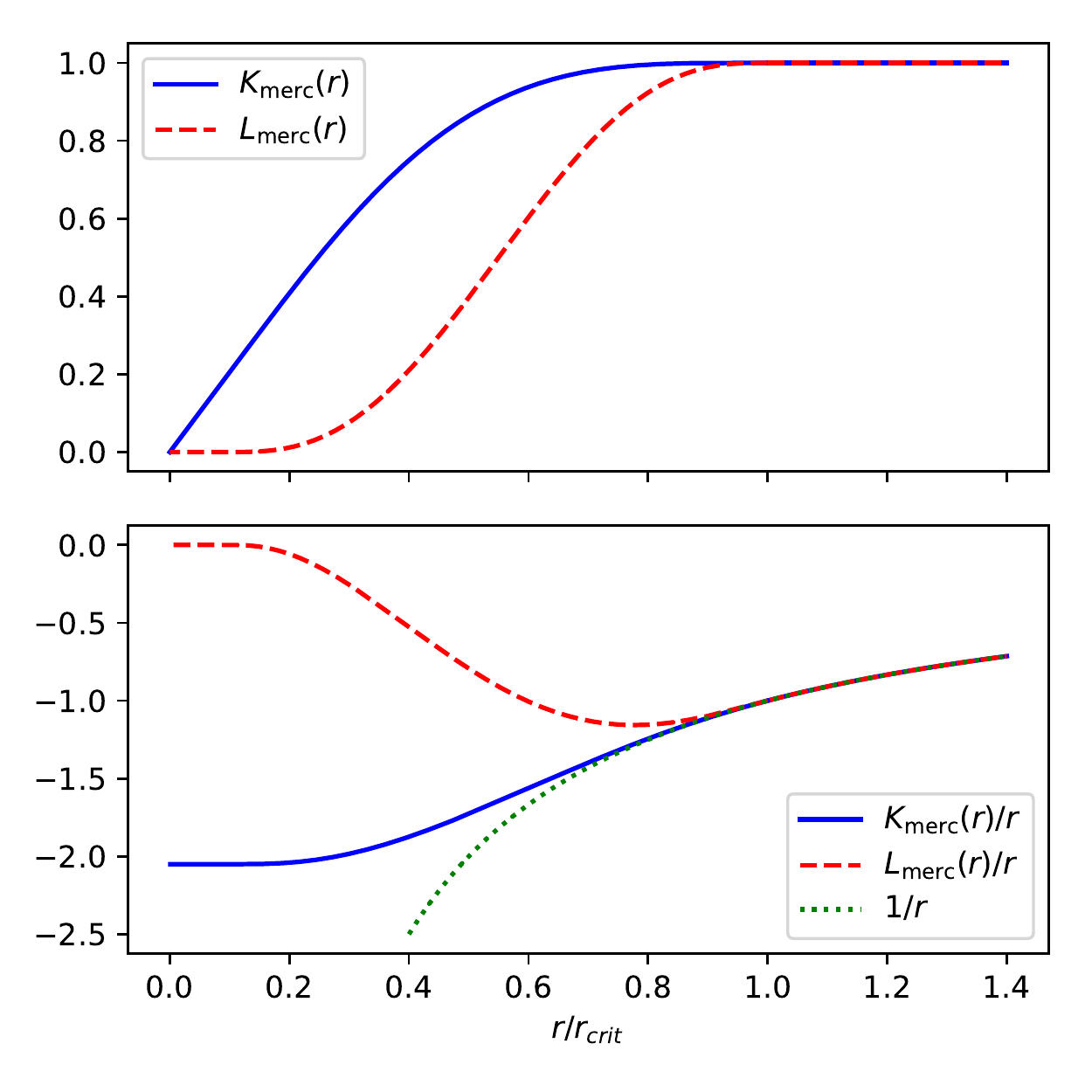}}
    \caption{Top panel: the Hamiltonian switching function $K_{\rm merc}$ and the corresponding force switching function $L_{\rm merc}$.
    Note that Chambers (1999) discusses $L_{\rm merc}$ in the context of a Hamiltonian switching function, but the actual implementation in \mer uses $L_{\rm merc}$, a force switching function.
    Bottom panel: potentials in the interaction Hamiltonian if both $K_{\rm merc}$ and $L_{\rm merc}$ were used as Hamiltonian switching functions. $K_{\rm merc}/r$ is the potential corresponding to the actual implementation of \mer.}
    \label{fig:switch}
\end{figure}
To find out what the corresponding Hamiltonians are that this scheme is integrating, we need to solve the differential equation in Eq.~(\ref{eq:LK}) to find the Hamiltonian switching function~$K$, given the force switching function~$L$. 
The solution is a one parameter family due to a gauge symmetry. 
We choose the solution that is continuous by matching $K_{\rm merc}(r_{\rm crit})=1$.
We plot the functions $K_{\rm merc}$ and $L_{\rm merc}$ in the top panel of Fig.~\ref{fig:switch}.
Note that the function $K_{\rm merc}$ does not reach exactly $0$ at $r=0.1r_{\rm crit}$. 
This means that there is a finite contribution to the interaction Hamiltonian $H_{\rm I}$ for any choice of $r>0$ (see function $K_{\rm merc}(r)/r$ in the bottom panel of Fig.~\ref{fig:switch}).

This is inconsistent with the discussion of \cite{Chambers1999} which argues that the switching function keeps the interaction Hamiltonian small.
That argument would only be correct if the force switching function $L_{\rm merc}$ were used as a Hamiltonian switching function (in the \mer code $L_{\rm merc}$ is used as a force switching function).
As an illustration, we also plot the function $L_{\rm merc}(r)/r$ in the bottom panel of Fig.~\ref{fig:switch}.
This is a potential which now reaches $0$ at $r=0.1r_{\rm crit}$.
However, whereas using $L_{\rm merc}$ as the Hamiltonian switching function would lead to a valid symplectic switching integrator, it will not perform as well as an integrator using $K_{\rm merc}$. 
Looking at the bottom panel of Fig.~\ref{fig:switch}, this is because $L_{\rm merc}(r)/r$ has a local minima around $0.8 r_{\rm crit}$ leading to an unphysical repulsive force between particles in some intermediate regime.
One consequence of the force changing sign is that the differential equations corresponding to $L_{\rm merc}(r)/r$ are stiffer during the close encounter.
Note that in both cases, particles will feel no force from the potentials in the interaction Hamiltonian while $r<0.1 r_{\rm crit}$.

It is worth pointing out that for small $r$ the shape of the effective potential $K_{\rm merc}(r)/r$ is similar to that of potentials used in other $N$-body systems where one simply ignores the close range interactions by having a finite smoothing length~(similar to a kernel in smooth particle hydrodynamics).

We would like to stress that despite this inconsistency in the derivation of the equations of motion, the \mer algorithm is solving a Hamiltonian system that converges to the original system in the limit of $dt\rightarrow 0$.

For the remainder of this paper, we will refer to the force switching function $L(r)$ simply as \textit{the switching function}. 
The corresponding Hamiltonian switching function $K(r)$ can always be calculated by solving Eq.~(\ref{eq:LK}). 

\section{Switching functions}
We now explore different force switching functions and their effects on the accuracy of the integration.
Our motivation comes from the fact that the function used by \mer has been determined heuristically and, as we have shown above, the original justification is not correct.
It is thus not clear if one can improve the integration scheme by simply choosing a better switching function.
In this paper, we will present and discuss results for the following four switching functions. 
These are extreme cases which will allow us to explain the general ideas behind choosing switching functions, which can then be applied to any arbitrary switching function.

(i) We set $L(r)=1$. The integrator simply becomes the (non-switching) Wisdom-Holman integrator in democratic heliocentric coordinates.

(ii) We set $L(r)$ to the discontinuous Heaviside function with a jump at $r=r_{\rm crit}$. 

(iii) Same as (ii) but the value of $L$ is kept constant for the duration of the entire timestep. 
To do that we approximate the trajectories along straight lines for the duration of the timestep and set $L=0$ if the trajectories come closer than $r_{\rm crit}$, and $L=1$ otherwise. 
There are cases where this breaks time reversibility. 
We could avoid this by doing this procedure iteratively. 
However, in all tests that we have performed this does not seem to be an issue.
Further, note that this procedure formally makes the integrator phase space dependent in a similar way than adaptive timesteps do. 
We comment on whether this has any real world consequences in Sec.~\ref{sec:disc}.

(iv) We use the switching function $L_{\rm merc}(r)$ given in Eq.(\ref{eq:L}). Thus the integrator becomes the \mer integrator.

(v) Finally, we consider an infinitely differentiable switching function given by
\begin{eqnarray}
    L_{\rm inf}(r) &=& = \frac{f(y)}{f(y)+f(1-y)}
\end{eqnarray}
where $y$ is given by Eq.~(\ref{eq:y}) and $f(y)$ is
\begin{eqnarray}
    f(y) = \begin{cases} 0 & y\leq0\\
        e^{-1/x} &y>0
    \end{cases}\label{eq:f}.
\end{eqnarray}
The shape of the function $L_{\rm inf}$ is very similar to that of $L_{\rm merc}$ (shown in Fig.~\ref{fig:switch}).
However the function $L_{\rm inf}(r)$ and all its derivatives are smooth and continuous for all $r$, even at the boundaries.
Note however, that just as $L_{\rm merc}(r)$, $L_{\rm inf}(r)$ is not analytic, i.e. a Taylor series expansion will not converge globally.

\section{Test Setup and Mercurius}
We now look at one representative test case in which a close encounter occurs between two giant planets.
We work in units where $G=1$ and the stellar mass is~$m_0=1$. 
Both planets have a mass similar to that of Jupiter,~$m_1=m_2=10^{-3}$.
The inner planet is on an orbit with semi-major axis~$a_1=1$ and eccentricity~$e_1=0.1$. 
The outer planet is on a nearby orbit with~$a_2= 1.1$ and~$e=0.2$.
We use a timestep corresponding to~0.5\% of the inner planet's period.
We adjust the phases such that a close encounter occurs at $t\approx 0$ with a close approach distance of approximately~$5\cdot 10^{-5}$.

To run these simulations we use \mercurius, our own implementation of the \mercury integrator\footnote{In reference to and appreciation of \mercury, we call our implementation \mercurius (the Latin word for Mercury).}.
We here only give a short overview of the implementation as we plan to publish a detailed code description paper. 
The \mercurius integrator is freely available as part of the REBOUND integrator package.
Internally it uses the Kepler solver of the WHFast integrator \citep{ReinTamayo2015} and the \ias integrator \citep{ReinSpiegel2015} to evolve $H_{\rm K}$.
Using \mercurius allows us to easily experiment with different switching functions.
We can also call the individual operators manually which allows us to calculate the commutators numerically as described above. 
We also ran tests with the original \mer code and implemented wrapper functions to make the fortran code callable from python.
These functions allow us to call \mer's time-stepping functions directly from python, avoiding any potential issues that may arise due to coordinate transformations, unit conversions, time-stepping logic, or input/output files.
This might have been an issue in an earlier study looking at the symplecticity of \mer (see Appendix~\ref{appendix:symplecticity}). 
Aside from small differences at the floating point precision limit, if we use the same switching function as implemented in \mer, we find perfect agreement between \mer and \mercurius in all our results and therefore only show the results using \mercurius.

\begin{figure*}
    \centering
    \resizebox{\textwidth}{!}{\includegraphics[trim=0.1cm 1.6cm 5cm 3cm,clip]{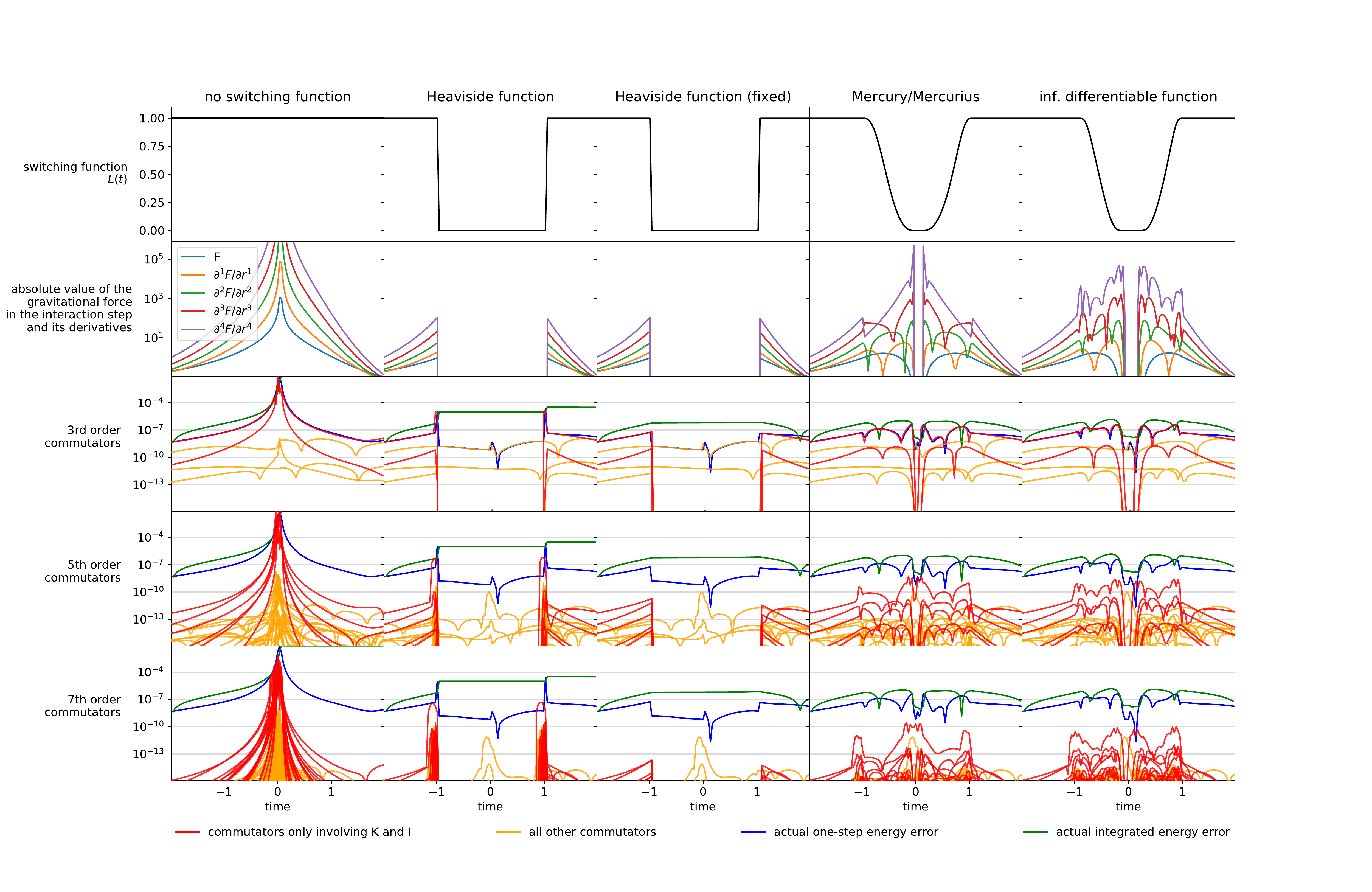}}
    \caption{
        Comparison of different switching functions used in a hybrid symplectic integrator during a close encounter of two giant planets.
        Each column corresponds to a different switching function. 
        The top row shows the value of the force switching function during the integration.
        The second row shows the absolute value of the gravitational force between the two planets in the interaction step.
        The bottom three rows show approximations of the third, fifth, and seventh order commutators that appear in the integrators' error terms.
        We work in a system of units where $G=1$. See text for more details. 
    \label{fig:main}
    }
\end{figure*}
\section{Results}

Figure~\ref{fig:main} summarizes our results for the test case and switching functions introduced above. 
The columns correspond to the different switching functions.
The horizontal axis of every panel corresponds to time, with the close encounter occurring at $t\approx 0$. 
The particles enter the critical distance $r_{\rm crit}$ at $t\approx -1$ and exit at it again at $t\approx 1$.
The top row shows the value of the switching function $L$ as a function of time.
The second row shows the absolute value of the gravitational force between the planets as calculated during the interaction step.
We also show the first four derivatives of the force.
The third, fourth, and fifth rows show approximations of the third, fifth, and seventh order commutators, respectively.
We colour commutators that only include $\widehat H_K$ and $\widehat H_I$ (i.e. not $\widehat H_J$) red, and all other commutators orange. 
We do not show commutators that are always zero (i.e. those only involving $\widehat H_I$ and $\widehat H_J$).
As a measure of the size of the commutators, we use the energy error as described above.
Note that the energy error does not account for phase errors.
Thus an energy error of zero does not imply a perfect solution.
The plots for the commutators also show the actual energy error occurred over one timestep step, as well as the integrated energy error, i.e. the energy error measured relative to the beginning of the simulation.

The first column shows the evolution for the standard WH integrator. 
The forces, its derivatives, and therefore all commutators get very large during the close encounter.
As a result, the integrated energy error goes off the chart. 
As expected, this integrator cannot accurately resolve the close encounter.
Note that the higher order commutators become comparable to the lower order commutators during the close encounter.
The physical interpretation for this is that timestep is too large to resolve the characteristic timescale (periastron passage of the close encounter).
Also note that the dominant commutators are those involving only~$\widehat H_K$ and~$\widehat H_I$.

The second column shows the evolution of using the Heaviside function as the switching function.
Note that the commutators which only involve $\widehat H_K$ and $\widehat H_I$ vanish when $L=0$.
This happens because $\widehat H_I$ becomes the identity operator, i.e. not having any effect, and it thus commutes with any other operator.
Commutators involving $\widehat H_K$ and $\widehat H_J$ do not vanish when $L=0$ and become the dominant contribution to the energy error during that time.
However, these are several orders of magnitude smaller and do not diverge during the close encounter unless the planets are very eccentric and close to periastron.
Note that there is a spike in most commutators exactly when $L$ transitions at $r_{\rm crit}$. 
This can be understood by imagining a scenario where the particles start out just outside of $r_{\rm crit}$. 
To calculate the commutator $\small[\widehat H_K,\widehat H_I\small]$, let us initially apply $\widehat H_I(dt)$.
When calculating the effect of this operator the value of $L$ will be $1$. 
If we then apply $\widehat H_K(dt)$ to the result, it might take the particles inside of $r_{\rm crit}$. 
Thus, if we then apply $\widehat H_I(-dt)$, the value of $L$ will be $0$.
Let us finally apply $\widehat H_K(-dt)$ again which might take the particles outside of $r_{\rm crit}$.
This sequence of operators corresponds to the commutator $\small[\widehat H_K,\widehat H_I\small]$ which is one of the nested commutators that appears at all orders in the BCH formula.
The above argument makes it clear that the commutator will be particularly large when the switching function changes a lot during a timestep.
This is the case for the Heaviside function, but it is important to note that for any other switching function that changes significantly during a timestep this is also the case, no matter if the switching function is smooth or has a discontinuity.
The consequence is that the actual energy error (both the single step error and the integrated error) jump up when the particles transition~$r_{\rm crit}$.
Despite this undesirable jump, the integrator is at least somewhat better at resolving the close encounter than the standard Wisdom-Holman integrator.
A physical argument can be given in terms of timescales again:  
if the switching function changes a lot during one timestep, we're effectively introducing a very short timescale that the integrator is unable to resolve.

We can avoid large changes of the switching function during the timestep, by simply calculating the value of $L$ once, at the beginning of the timestep, and then keeping it fixed. 
The results of this modification to the Heaviside switching function are shown in the third column.
Several features are identical to the standard way of using the Heaviside switching function (second column).
In particular, all commutators that only involve $\widehat H_K$ and $\widehat H_I$ still vanish when $L=0$.
However, we see that we have successfully removed the spikes in the commutators when $L$ changes.
The commutators involving $\widehat H_K$ and $\widehat H_I$ simply go to zero.
As a consequence the energy error does not increase when particles pass over the $r_{\rm crit}$ transition, despite the sudden change in the value of the switching function $L$. 
This can be explained following the same scenario described in the last paragraph. 
\textbf{This is a remarkable result:}
This integration scheme, effectively using a binary switching function, is significantly simpler to implement than those using a continuous switching function.
However, as one can see in the figure, it provides equal accuracy for one close encounter, at reduced computational complexity.
We further comment on this case below.

The fourth column shows the evolution using the standard \mer algorithm and its polynomial switching function. 
Once again all commutators that only involve $\widehat H_K$ and $\widehat H_I$ vanish when $L=0$.
However, in contrast to the integrators using the Heaviside function, this only happens for $r<0.1r_{\rm crit}$. 
Although these commutators are finite for $r>0.1r_{\rm crit}$ they are significantly smaller than in the case of the Wisdom-Holman integrator. 
Note that the third order commutators are smooth.
This is because the specific choice of switching function ensures that the forces and its first two derivatives are continuous. 
These derivatives appear in the Poisson bracket and therefore in the commutator.
However looking at the higher order commutators, one can see the effect of non-continuous higher derivates due to the finite differentiable switching function $L_{\rm merc}$. 
Fortunately, these commutators are significantly smaller than the third order commutators and therefore do not contribute much to the total energy error. 
Remember that the plots only show approximations of the commutators.
This becomes apparent in the 7th order commutators where the spikes near the discontinuities of the force derivatives at $r_{\rm crit}$ are somewhat smeared out. 

The fifth column shows the evolution using an infinitely differentiable switching function. 
Here, all derivatives of the force, and therefore all Poisson brackets, are continuous and smooth. 
Because of the smoothness requirement, the higher order derivatives of the force become highly oscillatory.
This is a generic feature of any infinitely differentiable switching function and not specific to the function we chose. 
The commutators look similar to those of the \mer algorithm but one significant difference is that the higher order commutators become larger and more oscillatory. 
As in the previous case this is because these commutators are directly related to the higher order derivatives of the forces. 
For moderate timesteps like the one we've chosen in this example, the energy error of the simulation is still dominated by the third order commutators. 
However, for larger timesteps, there can be cases where the higher order commutators will dominate the error.
We can think of this in physical terms once again.
The highly oscillatory derivatives of the force corresponds to very short timescales. 
Because we are using a finite and fixed timestep, these timescales can at some point be no longer accurately resolved, leading to larger errors.
The appearance of these small timescales is a purely numerical artefact \citep[somewhat related to timestep resonances discussed by][]{Rauch1999}.
Mathematically speaking, the differential equations corresponding to the higher order commutators become very stiff.
Note that in the standard Wisdom-Holman integrator (left panel), the forces get large, but do not oscillate.

\section{Conclusions}
\label{sec:disc}
We revisited the motivation behind hybrid symplectic integrators and the choice of switching functions.
We found that the derivation of equations of motion by \cite{Chambers1999} omits derivative terms. 
As we showed, it turns out that despite this inconsistency the \mer integrator is a switching integrator.
However, it effectively uses a very different switching function than the one described by \cite{Chambers1999}.

Motivated by the somewhat arbitrary choice of switching function in \mer, we explored a wide range of different switching functions to see if it is possible to improve the accuracy of this kind of integrators any further.
In particular, we presented results of two extreme cases: one infinitely differentiable function, and the Heaviside function.
In summary, we found that the smoothness alone is not a good criterion for choosing a switching function.

Our results show that an infinitely differentiable switching function does not perform much better over one close encounter than the polynomial function used by \mer.
We attribute this to two reasons. 
First, the beneficial effects of a smoother function only show up in higher order commutators which are several orders of magnitude smaller than the dominant third order commutators. 
Second, the higher order derivates of an infinitely differentiable switching function necessarily become highly oscillatory. 
The integrator cannot resolve the associated small timescales and thus behaves no better than an integrator involving finite differentiable functions.

We show that using a Heaviside switching function can have surprisingly good properties if the value of the switching function does not change during the timestep.
The resulting scheme is significantly simpler than the \mer scheme but achieves the same accuracy in the test case of a single close encounter presented.
What still needs to be studied in greater detail is the long term evolution in systems with repeated close encounters.
We ran additional tests of 1000~highly collisional planetary systems and found that the integrator using the Heaviside function performs on average as well as \mer.
We will discuss the implementation of this new integrator, as well as the implementation of the \mercurius integrator, in an upcoming code description paper.

In practice one is mostly interested in how accurate an integrator can resolve a close encounter.
In any numerical simulation, one can calculate a precise answer to this question using any arbitrary metric one thinks is approriate. 
A mathematical (or maybe even somehwat philosophical) question is whether these integrators we discussed in this paper are symplectic or not.
The authors of this paper were not able to agree on the answer to this question. 
We will therefore leave the final answer up to the reader, but comment on a few aspects of this question.

First, it is important to keep in mind that we very quickly reach the limits of finite floating point precision in systems with close encounters\footnote{Imagine a scenario where two planets on orbits with~$a\approx 1$ have a close encounter with a minimal encounter distance of one Earth radius,~$\approx10^{-5}$. Using double floating point precision, we only have~$\approx 11$ decimal digits to work with when representing the planets' positions in heliocentric coordinates.
When calculating any derivative, which is required for checking the symplecticity, we further loose roughly half of the significant digits.}.
One could argue that we should not talk about concepts of differential geometry such as symplecticity at all if we work with floating point numbers because derivatives formally do not even exist.
We run simulations on a discrete phase space where only finite differences exist.

Second, note that all switching functions we used are non-analytic.
That is not a coincidence but a requirement if we want the integrators to fall back to the Wisdom-Holman integrator whenever particles are far away from each other. 
As a consequence, the Taylor series of the switching function, and therefore the Poisson Brackets in the BCH formula will not converge to the correct solution globally. 

Third, when choosing a switching function, differentiability is not the right criteria to consider.
If the mathematical concept of differtiability were an important requirement for a switching function, we could simply come up with a new infinitely differentiable function that is arbitrarily close to the original non-smooth function.
We have effectivly done that for $L_{\rm merc}$ by introducing $L_{\rm inf}$, which did not lead to improvements in any of our tests.
What matters is how much the switching function changes during a (finite) timestep.
If it changes a lot, then the corresponding differential equations become stiff and the errors large.
What also matters is that the switching function is continous to ensure the existance and uniquness of a solution\footnote{This only matters when calculating the solution of the $H_K$ step. 
It does not matter for the solution of the $H_I$ step because the switching function is constant.} \citep{2019arXiv190203684H}.
Note that integrator with the Heaviside function completely changes the nature of the algorithm and can among other things break time reversibility.
This might be important for some long term integrations, but not for simulations where encounters are rare (for example for planetary system which go unstable), or if physical collisions break time reversibility anyway.

Fourth, all integrators will converge to the correct solution in the limit of the timestep going to zero.
We thus argue that the simplest and most reliable way to test that a given integrator gives reliable answers is to change the timestep and watch for changes in the results, i.e. a classical convergence test.

\section*{Acknowledgments}
We thank the referee John Chambers for providing us with helpful comments which allowed us to improve and clarify the manuscript.
We thank Scott Tremaine for helpful discussions at various stages of this project.
This research has been supported by the NSERC Discovery Grant RGPIN-2014-04553, the 2018 Fields Undergraduate Summer Research Program at the Fields Institute for Research in Mathematical Sciences in Toronto, and the Centre for Planetary Sciences at the University of Toronto Scarborough.
Support for this work was also provided by NASA through the NASA Hubble Fellowship grant HST-HF2-51423.001-A awarded by the Space Telescope Science Institute, which is operated by the Association of Universities for Research in Astronomy, Inc., for NASA, under contract NAS5-26555.
This research was made possible by the open-source projects 
\texttt{Jupyter} \citep{jupyter}, \texttt{iPython} \citep{ipython}, 
and \texttt{matplotlib} \citep{matplotlib, matplotlib2}.

\bibliography{full}

\appendix

\section{Democratic Heliocentric Coordinates}
\label{appendix:coord}
The hybrid integration schemes discussed in this paper make use of democratic heliocentric coordinates \citep{Duncan1998}.
Let us assume we have one star and an additional $N$ planetary bodies with Cartesian coordinates $\mathbf q_i$ and canonical momenta $\mathbf p_i$ in some inertial frame.
Then, the democratic heliocentric coordinates are defines as:
\begin{eqnarray}
    \mathbf Q_i = 
        \begin{cases}
          \mathbf q_i - \mathbf q_0  & i \neq 0\\
          \frac 1 M \sum_{j=0}^{N-1} m_j \mathbf q_j & i = 0.
        \end{cases}
\end{eqnarray}
Here, $M=\sum_{j=0}^{N-1}m_j$ is the total mass of the system. The corresponding canonical momenta are
\begin{eqnarray}
    \mathbf P_i = 
        \begin{cases}
          \mathbf p_i - \frac{m_i}M \sum_{j=0}^{N-1} \mathbf p_j  & i \neq 0\\
           \sum_{j=0}^{N-1} \mathbf p_j & i = 0.
        \end{cases}
\end{eqnarray}

\section{Symplecticity of Mercury}
\label{appendix:symplecticity}
While working with \mer and \mercurius we also looked at measuring the symplecticity error.
To do that we follow \cite{Hernandez2016} in calculating the Jacobian using a finite difference approach. 
This is non-trivial because one quickly runs into issues of finite floating point precision. 
The results of \cite{Hernandez2016} suggest that the \mer algorithm might be non-symplectic for their binary planet test case.
To investigate this issue, we implemented wrapper functions to directly call the \mer subroutines which evolve the state by one timestep, but ignore all the input, output and other bookkeeping login in the \mer package.
Our results with this method differ from those obtained by \cite{Hernandez2016} and indicate that \mer is symplectic (whether we should really call the scheme symplectic or not, becomes a somewhat philosophical rather than practical question, see discussion).

We attribute this discrepancy to some non-symplectic operations that the \mer package applies to the state vector before and after the actual integration (e.g. unit conversions and coordinate transformation), and a loss of precision when using ASCII files as input and output files.
We also ran tests with our own implementation of the algorithm, \mercurius, and confirm the results obtained with \mer.

\end{document}